\documentclass[aps,pra,twocolumn,showpacs,10pt]{revtex4}
\usepackage{graphicx}
\usepackage{amssymb}
\usepackage{amsmath}
\begin{document}
\title{Positive-P phase space method simulation in superradiant emission from a cascade atomic ensemble}
\author{H. H. Jen}
\affiliation{Physics Department, National Tsing Hua University, Hsinchu 300, Taiwan, R. O. C.}
\date{\today}

\begin{abstract}The superradiant emission properties from an atomic ensemble with cascade level configuration is numerically simulated.  The correlated spontaneous emissions (signal then idler fields) are purely stochastic processes which are initiated by quantum fluctuations.  We utilize the positive-P phase space method to investigate the dynamics of the atoms and counter-propagating emissions.  The light field intensities are calculated, and the signal-idler correlation function is studied for different optical depths of the atomic ensemble.  Shorter correlation time scale for a denser atomic ensemble implies a broader spectral window needed to store or retrieve the idler pulse.
\end{abstract}
\maketitle
\section{Introduction}
A quantum communication network based on the distribution and sharing of
entangled states is potentially secure to eavesdropping and is therefore of
great practical interest \cite{QI,cryp,QI2}. \ A protocol for the realization
of such a long distance system, known as the quantum repeater, was proposed by
Briegel \textit{et al}. \cite{repeater,Dur}. \ A quantum repeater based on the
use of atomic ensembles as memory elements, distributed over the network, was
subsequently suggested by Duan, Lukin, Cirac and Zoller \cite{dlcz}. \ The
storage of information in the atomic ensembles involves the Raman scattering
of an incident light beam from ground state atoms with the emission of a
signal photon. \ The photon is correlated with the creation of a phased,
ground-state, coherent excitation of the atomic ensemble. \ The information
may be retrieved by a reverse Raman scattering process, sending the excitation
back to the initial atomic ground state and generating an idler photon
directionally correlated with the signal photon
\cite{qubit,chou,vuletic,collective,store,collective2,single2,pan,kimble}.
\ In the alkali gases, the signal and the idler field wavelengths are in the
near-infrared spectral region. \ This presents a wavelength mismatch with
telecommunication wavelength optical fiber, which has a transmission window at
longer wavelengths\ (1.1-1.6 um). \ It is this mismatch that motivates the
search for alternative processes that can generate telecom wavelength photons
correlated with atomic spin waves \cite{telecom}. \ 

This motivates the research presented in this article where we study
multi-level atomic schemes in which the transition between the excited states
is resonant with a telecom wavelength light field \cite{telecom}. \ The basic
problem is to harness the absorption and the emission of telecom photons while
preserving quantum correlations between the atoms, which store information and
the photons that carry along the optical fiber channel of the network. 

It is not common to have a telecom ground state transition in atomic gases
except for rare earth elements \cite{erbium,dysprosium} or in an erbium-doped
crystal \cite{solid}. \ However, a telecom wavelength (signal) can be
generated from transitions between excited levels in the alkali metals
\cite{telecom,radaev}. \ 

The ladder configuration of atomic levels provides a source for telecom
photons (signal) from the upper atomic transition. \ For rubidium and cesium
atoms, the signal field has the range around 1.3-1.5 $\mu$m that can be
coupled to an optical fiber and transmitted to a remote location. \ Cascade
emission may result in pairs of photons, the signal entangled with the
subsequently emitted infrared photon (idler) from the lower atomic transition.
\ Entangled signal and idler photons were generated from a phase-matched
four-wave mixing configuration in a cold, optically thick $^{85}$Rb ensemble
\cite{telecom}. \ This correlated two-photon source is potentially useful as
the signal field has telecom wavelength.

The temporal emission characteristics of the idler field, generated on the
lower arm of the cascade transition, were observed in measurements of the
joint signal-idler correlation function. \ The idler decay time was shorter
than the natural atomic decay time and dependent on optical thickness in a way
reminiscent of superradiance \cite{Dicke,Stephen,Lehm,mu,OC:Mandel}.

The spontaneous emission from an optically dense atomic ensemble is a
many-body problem due to the radiative coupling between atoms. \ This coupling
is responsible for the phenomenon of superradiance firstly discussed by Dicke
\cite{Dicke} in 1954.

Since then, this collective emission has been extensively studied in two atom
systems indicating a dipole-dipole interaction \cite{Stephen,Lehm}, in the
totally inverted N atom systems \cite{stehle,Tallet}, and in the extended
atomic ensemble \cite{mu}. \ The emission intensity has been investigated
using the master equation approach \cite{master,Bon,Bon1} and with
Maxwell-Bloch equations \cite{Bon2,Feld}. \ A useful summary and review of
superradiance can be found in the reference \cite{Gross,phase}. \ Recent
approaches to superradiance include the quantum trajectory method
\cite{trajectory,eig1} and the quantum correction method \cite{Fleischhauer99}%
. \ 

In the limit of single atomic excitation, superradiant emission
characteristics have been discussed in the reference \cite{Eberly} and
\cite{Scully}. \ For a singly excited system, the basis set reduces to N
rather than $2^{N}$ states. \ Radiative phenomena have been investigated using
dynamical methods \cite{kurizki,Scully2,eig2} and by the numerical solution of
an eigenvalue problem \cite{Friedberg08a,
Svidzinsky08,Friedberg08b,Friedberg08c}. \ A collective frequency shift
\cite{Arecchi, Morawitz} can be significant at a high atomic density
\cite{Scully09} and has been observed recently in an experiment where atoms
are resonant with a planar cavity \cite{supershift}.

To account for multiple atomic excitations in the signal-idler emission from a
cascade atomic ensemble, the Schr\"{o}dinger's equation approach becomes
cumbersome. \ An alternative theory of c-number Langevin equations is suitable for solution by stochastic simulations.  

Langevin equations were initially derived to describe Brownian motion \cite{SM:Gardiner}. \ A fluctuating force is used to represent the random
impacts of the environment on the Brownian particle. \ A given realization of
the Langevin equation involves a trajectory perturbed by the random force.
\ Ensemble averaging such trajectories provides a natural and direct way to
investigate the dynamics of the stochastic variables. \ 

\ An essential element in the stochastic simulations is a proper
characterization of the Langevin noises. \ These represent the quantum
fluctuations responsible for the initiation of the spontaneous emission from
the inverted \cite{Feld,Haake1,Haake2,Polder79}, or pumped atomic system
\cite{Chiao88,Chiao95} as in our case.\ \ 

The positive-P phase space method \cite{QN:Gardiner,
quantization,Smith88,Smith0,Smith1,Boyd89,Drummond91} is employed to derive
the Fokker-Planck equations that lead directly to the c-number Langevin
equations. \ The classical noise correlation functions, equivalently diffusion
coefficients, are alternatively\ confirmed by use of the Einstein relations \cite{LP:Sargent, QO:Scully, Fleischhauer94}. 
\ The c-number Langevin equations correspond to
Ito-type stochastic differential equations that may be simulated numerically.
\ The noise correlations can be represented either by using a square
\cite{Carmichael86} or a non-square "square root" diffusion matrix
\cite{Smith1}. \ The approach enables us to calculate normally-ordered
quantities, signal-idler field intensities, and the second-order correlation
function. \ The numerical approach involves a semi-implicit difference
algorithm and shooting method \cite{numerical} to integrate the stochastic
"Maxwell-Bloch" equations.

Recently a new positive-P phase space method involving a stochastic gauge
function \cite{Drummond02} has been developed. \ This approach has an improved
treatment of sampling errors and boundary errors in the treatment of quantum
anharmonic oscillators \cite{Drummond01,Collett01}. \ It has also been applied
to a many-body system of bosons \cite{Drummond03} and fermions
\cite{Drummond06}. \ In this paper, we follow the traditional positive-P
representation method \cite{drummond80}.\ 

The remainder of this paper is organized as follows.  In section II, we show the formalism of positive P-representation, and demonstrate the stochastic differential equations of cascade emission (signal and idler) from an atomic ensemble.  In section III we solve numerically for the dynamics of the atoms and counter-propagating signal and idler fields in a positive P-representation.  We present results of signal and idler field intensities, and the signal-idler second order correlation function for different optical depths of the atomic ensemble.  Section IV presents our discussions and conclusions.  
In the appendix, we show the details in the derivations of c-number Langevin equations that are the foundation for numerical approaches of the cascade
emission. \ In Appendix A, we formulate the Hamiltonian, and derive the Fokker-Planck equations by characteristic functions \cite{LT:Haken} in positive P-representation.  Then corresponding c-number Langevin equations are derived, and the noise correlations are found from the diffusion coefficients in Fokker-Planck equations as shown in Appendix B. \ 

\section{Theory of Cascade emission}

The phase space methods \cite{QN:Gardiner} that mainly include P-, Q-, and Wigner (W) representations are techniques of using classical analogues to study quantum systems, especially harmonic oscillators. \ The eigenstate of harmonic oscillator is a coherent state that provides the basis expansion to construct various representations. \ P and Q-representation are associated respectively with evaluations of normal and anti-normal order correlations of creation and destruction operators. \ W-representation is invented for the
purpose of describing symmetrically ordered creation and destruction operators. \ Since P-representation describes normally ordered quantities that are relevant in experiments, we are interested in investigating one class of generalized P-representations, the positive P-representation that has semi-definite property in the diffusion process, which is important in describing quantum noise systems.

Positive-P representation \cite{QO:Walls, drummond80} is an extension to Glauber-Sudarshan P-representation that uses coherent state ($|\alpha\rangle$) as a basis expansion of density operator $\rho$. \ In terms of diagonal coherent states with a quasi-probability distribution, $P(\alpha,\alpha^{\ast})$, a density operator in P-representation is
\begin{equation}
\rho=\int_{D}|\alpha\rangle\langle\alpha|P(\alpha,\alpha^{\ast})d^{2}\alpha,
\end{equation}
where $D$ represents the integration domain. \ The normalization condition of $\rho,$ which is Tr\{$\rho$\}$=1,$ indicates the normalization for $P$ as well, $\int_{D}P(\alpha,\alpha^{\ast})d^{2}\alpha=1$. \ 

Positive P-representation uses a non-diagonal coherent state expansion and the density operator can be expressed as%
\begin{equation}
\rho=\int_{D}\Lambda(\alpha,\beta)P(\alpha,\beta)d\mu(\alpha,\beta),
\end{equation}
where%

\begin{equation}
d\mu(\alpha,\beta)=d^{2}\alpha d^{2}\beta\text{ and }\Lambda(\alpha,\beta)=\frac{|\alpha\rangle\langle\beta^{\ast}|}{\langle\beta^{\ast}%
|\alpha\rangle},
\end{equation}
and $\langle\beta^{\ast}|\alpha\rangle$ in non-diagonal projection operators, $\Lambda(\alpha,\beta),$ makes sure of the normalization condition in distribution function, $P(\alpha,\beta).$

Any normally ordered observable can be deduced from the distribution function
$P(\alpha,\beta)$ that
\begin{equation}
\langle(a^{\dag})^{m}a^{n}\rangle=\int_{D}\beta^{m}\alpha^{n}P(\alpha,\beta)d\mu(\alpha,\beta).
\end{equation}

A characteristic function $\chi_{p}(\lambda_{\alpha},\lambda_{\beta})$ (Fourier-transformed distribution function in Glauber-Sudarshan
P-representation but now is extended into a larger dimension) can help formulate distribution function, which is%
\begin{equation}
\chi_{p}(\lambda_{\alpha},\lambda_{\beta})=\int_{D}e^{i\lambda_{\alpha}\alpha+i\lambda_{\beta}\beta}P(\alpha,\beta)d\mu(\alpha,\beta).
\end{equation}
It is calculated from a normally ordered exponential operator $E(\lambda),$
\begin{equation}
\chi_{p}(\lambda_{\alpha},\lambda_{\beta})=\text{Tr\{}\rho E(\lambda)\text{\},
}E(\lambda)=e^{i\lambda_{\beta}a^{\dagger}}e^{i\lambda_{\alpha}a}.
\end{equation}

Then a Fokker-Planck equation can be derived from the time derivative of characteristic function,%

\begin{equation}
\frac{\partial\chi_{p}}{\partial t}=\frac{\partial}{\partial t}\text{Tr\{}\rho E(\lambda)\text{\}=Tr\{}\frac{\partial\rho}{\partial t}E(\lambda)\text{\}}%
\end{equation}
by Liouville equations,%
\begin{equation}
\frac{\partial\rho}{\partial t}=\frac{1}{i\hbar}[H,\rho].
\end{equation}

\begin{figure}
[ptb]
\begin{center}
\includegraphics[
natheight=7.499600in,
natwidth=9.999800in,
height=2.1037in,
width=3.0682in]%
{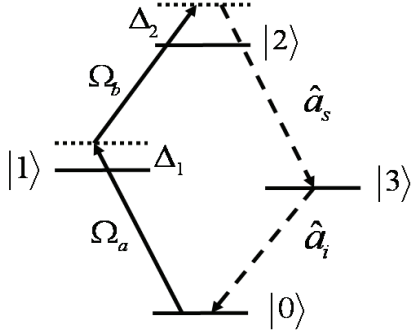}%
\caption{Four-level atomic ensemble interacting with two driving lasers
(solid) with Rabi frequencies $\Omega_{a}$ and $\Omega_{b}.$ \ Signal and
idler fields are labelled by $\hat{a}_{s}$ and $\hat{a}_{i},$ respectively and
$\Delta_{1}$ and $\Delta_{2}$ are one and two-photon laser detunings. }%
\label{four}%
\end{center}
\end{figure}

In laser theory \cite{LT:Haken}, a P-representation method is extended to describe atomic and atom-field interaction systems. \ When a large number of atoms is considered, which is indeed the case of the actual laser, a macroscopic variable can be defined. \ Then a generalized Fokker-Planck equation can be derived from characteristic functions by neglecting higher order terms that are proportional to the inverse of number of atoms. \ It is similar to our case when we solve light-matter interactions in an atomic ensemble that the large number cuts off the higher order terms in characteristic functions.

We consider $N$ cold atoms that are initially prepared in the ground state
interacting with four independent electromagnetic fields.\ \ As shown in
Fig.\ref{four}, two driving lasers (of Rabi frequencies $\Omega_{a}$ and
$\Omega_{b}$) excite a ladder configuration $|0\rangle\rightarrow
|1\rangle\rightarrow|2\rangle.$ \ Two quantum fields, signal $\hat{a}_{s}$ and
idler $\hat{a}_{i},$ are generated spontaneously.  We note that the spontaneous 
emission from the cascade driving scheme is a stochastic process due to the quantum fluctuations, 
unlike the diamond configuration where quantum noise can be neglected \cite{conversion,thesis}.

The complete derivation of the c-number Langevin equations for cascade
emission from the four-level atomic ensemble is described in Appendix A and B. \ After setting up the Hamiltonian, we follow the standard
procedure to construct the characteristic functions \cite{LT:Haken} in
Appendix A using the positive-P representation \cite{QN:Gardiner}. \ In
Appendix B.1, the Fokker-Planck equation is found by directly Fourier
transforming the characteristic functions, and making a $1/N_{z}$ expansion.\ 

\ Finally the Ito stochastic differential equations are written down from
inspection of the first-order derivative (drift term) and second-order
derivative (diffusion term) in the Fokker-Planck equation. \ The equations are
then written in dimensionless form by introducing the Arecchi-Courtens
cooperation units \cite{scale} in Appendix B.2. \ From Eq. (\ref{bloch2})
and the field equations that follow, these c-number Langevin equations in a
co-moving frame are,%
\begin{widetext}
\begin{align}
\frac{\partial}{\partial\tau}\pi_{01}  & =(i\Delta_{1}-\frac{\gamma_{01}}%
{2})\pi_{01}+i\Omega_{a}(\pi_{00}-\pi_{11})+i\Omega_{b}^{\ast}\pi_{02}%
-i\pi_{13}^{\dag}E_{i}^{+}+\mathcal{F}_{01}\text{ (I),}\nonumber\\
\frac{\partial}{\partial\tau}\pi_{12}  & =i(\Delta_{2}-\Delta_{1}%
+i\frac{\gamma_{01}+\gamma_{2}}{2})\pi_{12}-i\Omega_{a}^{\ast}\pi_{02}%
+i\Omega_{b}(\pi_{11}-\pi_{22})+i\pi_{13}E_{s}^{+}e^{-i\Delta kz} +\mathcal{F}_{12},\nonumber\\
\frac{\partial}{\partial\tau}\pi_{02}  & =(i\Delta_{2}-\frac{\gamma_{2}}%
{2})\pi_{02}-i\Omega_{a}\pi_{12}+i\Omega_{b}\pi_{01}+i\pi_{03}E_{s}%
^{+}e^{-i\Delta kz}-i\pi_{32}E_{i}^{+}+\mathcal{F}_{02},\nonumber\\
\frac{\partial}{\partial\tau}\pi_{11}  & =-\gamma_{01}\pi_{11}+\gamma_{12}%
\pi_{22}+i\Omega_{a}\pi_{01}^{\dag}-i\Omega_{a}^{\ast}\pi_{01}-i\Omega_{b}%
\pi_{12}^{\dag}+i\Omega_{b}^{\ast}\pi_{12}+\mathcal{F}_{11},\nonumber
\end{align}\begin{align}
\frac{\partial}{\partial\tau}\pi_{22}  & =-\gamma_{2}\pi_{22}+i\Omega_{b}%
\pi_{12}^{\dag}-i\Omega_{b}^{\ast}\pi_{12}+i\pi_{32}^{\dag}E_{s}%
^{+}e^{-i\Delta kz}-i\pi_{32}E_{s}^{-}e^{i\Delta kz}+\mathcal{F}%
_{22},\nonumber\\
\frac{\partial}{\partial\tau}\pi_{33}  & =-\gamma_{03}\pi_{33}+\gamma_{32}%
\pi_{22}-i\pi_{32}^{\dag}E_{s}^{+}e^{-i\Delta kz}+i\pi_{32}E_{s}^{-}e^{i\Delta
kz}+i\pi_{03}^{\dag}E_{i}^{+}-i\pi_{03}E_{i}^{-} +\mathcal{F}_{33},\nonumber\\
\frac{\partial}{\partial\tau}\pi_{13}  & =-(i\Delta_{1}+\frac{\gamma
_{01}+\gamma_{03}}{2})\pi_{13}-i\Omega_{a}^{\ast}\pi_{03}-i\Omega_{b}\pi
_{32}^{\dag}+i\pi_{12}E_{s}^{-}e^{i\Delta kz}+i\pi_{01}^{\dag}E_{i}%
^{+} +\mathcal{F}_{13},\nonumber\\
\frac{\partial}{\partial\tau}\pi_{03}  & =-\frac{\gamma_{03}}{2}\pi
_{03}-i\Omega_{a}\pi_{13}+i\pi_{02}E_{s}^{-}e^{i\Delta kz}+i(\pi_{00}-\pi
_{33})E_{i}^{+}+\mathcal{F}_{03},\nonumber\\
\frac{\partial}{\partial\tau}\pi_{32}  & =i\Delta_{2}-\frac{\gamma_{03}%
+\gamma_{2}}{2}\pi_{32}+i\Omega_{b}\pi_{13}^{\dag}-i(\pi_{22}-\pi_{33}%
)E_{s}^{+}e^{-i\Delta kz}-i\pi_{02}E_{i}^{-}+\mathcal{F}_{32},\nonumber\\
\frac{\partial}{\partial z}E_{s}^{+}  & =-i\pi_{32}e^{i\Delta kz}\frac
{|g_{s}|^{2}}{|g_{i}|^{2}}-\mathcal{F}_{s},\text{ }\frac{\partial}{\partial
z}E_{i}^{+}=i\pi_{03}+\mathcal{F}_{i}, \label{bloch3}%
\end{align}
\end{widetext}
where (I) stands for Ito type SDE. \ $\pi_{ij}$ is the stochastic variable
that corresponds to the atomic populations of state $|i\rangle$ when $i=j$ and
to atomic coherence when $i\neq j$, and $\mathcal{F}_{ij}$ are c-number
Langevin noises. \ The remaining equations of motion, which close the set, can
be found by replacing the above classical variables, $\pi_{jk}^{\ast
}\rightarrow\pi_{jk}^{\dag},$ $(\pi_{jk}^{\dag})^{\ast}\rightarrow\pi_{jk},$
$(E_{s,i}^{+})^{\ast}\rightarrow E_{s,i}^{-},$ $(E_{s,i}^{-})^{\ast
}\rightarrow E_{s,i}^{+}$ , and $\mathcal{F}_{jk}^{\ast}\rightarrow
\mathcal{F}_{jk}^{\dag}$.\ \ Note that the atomic populations satisfy
$\pi_{jj}^{\ast}=\pi_{jj}.$ \ The superscripts, dagger ($\dag$) for atomic
variables and ($-$) for field variables, denote the independent variables,
which is a feature of the positive-P representation: there are double
dimension spaces for each variable. \ These variables are complex conjugate to
each other when ensemble averages are taken, for example $\left\langle
\pi_{jk}\right\rangle =\left\langle \pi_{jk}^{\dag}\right\rangle ^{\ast}$ and
$\left\langle E_{s,i}^{+}\right\rangle =\left\langle E_{s,i}^{-}\right\rangle
^{\ast}.$ \ The doubled spaces allow the variables to explore trajectories
outside the classical phase space.

Before going further to discuss the numerical solution of the SDE, we point
out that the diffusion matrix elements have been computed using Fokker-Planck
equations and by the Einstein relations discussed in Appendix B.2. \ This
provides the important check on the lengthy derivations of the diffusion
matrix elements we need for the simulations.

The next step is to find expressions for the Langevin noises in terms of a non-square matrix $B$
\cite{QO:Walls,Smith1}. \ The matrix $B$ is used to construct the symmetric
diffusion matrix $D(\alpha)=B(\alpha)B^{T}(\alpha)$ for a Ito SDE,%

\begin{equation}
dx_{t}^{i}=A_{i}(t,\overrightarrow{x_{t}})dt+\sum\limits_{j}B_{ij}%
(t,\overrightarrow{x_{t}})dW_{t}^{j}(t)\text{ \ (I)}\label{Ito}%
\end{equation}
where $\xi_{i}dt=dW_{t}^{i}(t)$ (Wiener process) and $\left\langle \xi
_{i}(t)\xi_{j}(t^{\prime})\right\rangle =\delta_{ij}\delta(t-t^{\prime}).$
\ Note that $B\rightarrow BS,$ where $S$ is an orthogonal matrix ($SS^{T}=I$),
leaves $D$\ unchanged, so $B$ is not unique. \ We could also construct a
square matrix representation $B$ \cite{QN:Gardiner,SM:Gardiner,Carmichael86}.
\ This involves a procedure of matrix decomposition into a product of lower
and upper triangular matrix factors. \ A Cholesky decomposition can be used to
determine the $B$ matrix elements successively row by row. \ The downside of
this procedure is that the $B$ matrix elements must be differentiated in
converting the Ito SDE to its equivalent Stratonovich form for numerical solution.

The Stratonovich SDE is necessary for the stability and the convergence of
semi-implicit methods. \ Because of the analytic difficulties in transforming
to the Stratonovich form, we use instead the non-square form of $B$
\cite{Smith1}. \ 

In this case a typical $B$ matrix element is a sum of terms, each one of which
is a product of the square root of a diffusion matrix element with a unit
strength real (if the diffusion matrix element is diagonal) or complex (if the
diffusion matrix element is off-diagonal) Gaussian unit white noise. \ It is
straightforward to check that a $B$ matrix constructed in this way reproduces
the required diffusion matrix $D=BB^{T}$.

As pointed out in the reference \cite{Drummond91}, the transverse
dipole-dipole interaction can be neglected and nonparaxial spontaneous decay
rate can be accounted for by a single atom decay rate if the atomic
density is not too high. \ We are interested here in conditions where the
ensemble length $L$ is significant and propagation effects are non-negligible,
and the average distance between atoms $d=\sqrt[3]{V/N}$ is larger than the
transition wavelength $\lambda.$ \ The length scales satisfy $\lambda\lesssim
d\ll L,$ and we consider a pencil-like cylindrical atomic ensemble. \ The
paraxial or one-dimensional assumption for field propagation is then valid,
and the transverse dipole-dipole interaction is not important for the atomic
density we focus here.

The theory of cascade emission presented here provides the solid ground for simulations of 
fluctuations that initiate the radiation process in the atomic ensemble.  A proper way of treating 
fluctuations or noise correlations and formulating SDE requires an Ito form that is derived from the Fokker-Planck equation.  
An alternative but more straightforward approach by making quantum to classical correspondence in the quantum Langevin equation 
does not guarantee an Ito type SDE.  That is the reason we take the route of Fokker-Planck equation, and the coupled equations of Eq.(\ref{bloch3}) are the main results in this section. 

\section{Results for signal, idler intensities, and the second-order
correlation function}

There are several possible ways to integrate the differential equation
numerically. \ Three main categories of algorithm used are forward (explicit),
backward (implicit), and mid-point (semi-implicit) methods \cite{numerical}.  
The forward difference method, which Euler or Runge-Kutta methods utilizes, is
not guaranteed to\ converge in stochastic integrations \cite{xmds}. \ There it
is shown that the semi-implicit method \cite{semi} is more robust in
Stratonovich type SDE simulations \cite{Drummond91b}. \ More extensive studies
of the stability and convergence of SDE can be found in the reference
\cite{SDE:Kloeden}. \ The Stratonovich type SDE equivalent to the Ito type
equation (\ref{Ito}), is%
\begin{align}
dx_{t}^{i}  & =[A_{i}(t,\overrightarrow{x_{t}})-\frac{1}{2}\sum\limits_{j}%
\sum\limits_{k}B_{jk}(t,\overrightarrow{x_{t}})\frac{\partial}{\partial x^{j}%
}B_{ik}(t,\overrightarrow{x_{t}})]dt\nonumber\\
& +\sum\limits_{j}B_{ij}(t,\overrightarrow{x_{t}})dW_{t}^{j}\text{
\ (Stratonovich),}%
\end{align}
which has the same diffusion terms $B_{ij},$ but with modified drift terms.
\ This "correction" term arises from the different definitions of stochastic
integral in the Ito and Stratonovich calculus.

At the end of Appendix C.3, we derive the "correction" terms noted above. \ We then have 19 classical
variables including atomic populations, coherences, and two
counter-propagating cascade fields. \ With 64 diffusion matrix elements and an
associated 117 random numbers required to represent the instantaneous Langevin
noises, we are ready to solve the equations numerically using the robust
midpoint difference method.

The problem we encounter here involves counter-propagating field equations in
the space dimension and initial value type atomic equations in the time
dimension.  The counter-propagating field equations have a boundary condition specified at
each end of the medium. \ This is a two-point boundary value problem, and a
numerical approach to its solution, the shooting method \cite{numerical}, is used here.

Any normally-ordered quantity $\left\langle Q\right\rangle $ can be derived by
ensemble averages that $\left\langle Q\right\rangle =\sum_{i=1}^{R}Q_{i}/R $
where $Q_{i}$ is the result for each realization.

In this section, we present the second-order correlation function of
signal-idler fields, and their intensity profiles. $\ $We define the
intensities of signal and idler fields by \ %

\begin{equation}
I_{s}(t)=\left\langle E_{s}^{-}(t)E_{s}^{+}(t)\right\rangle ,\text{ }%
I_{i}(t)=\left\langle E_{i}^{-}(t)E_{i}^{+}(t)\right\rangle ,
\end{equation}
respectively, and the second-order signal-idler correlation function%

\begin{equation}
G_{s,i}(t,\tau)=\left\langle E_{s}^{-}(t)E_{i}^{-}(t+\tau)E_{i}^{+}%
(t+\tau)E_{s}^{+}(t)\right\rangle
\end{equation}
where $\tau$ is the delay time of the idler field with respect a reference
time $t$\ of the signal field. \ Since the correlation function is not
stationary \cite{QL:Loudon}, we choose $t$ as the time when $G_{s,i}$ is at
its maximum.

We consider a cigar shaped $^{85}$Rb ensemble of radius $0.25$ mm and $L=3$
mm. \ The operating conditions of the pump lasers are ($\Omega_{a},$
$\Omega_{b},$ $\Delta_{1},$ $\Delta_{2}$) $=$ ($0.4,$ $1,$ $1,$ $0$%
)$\gamma_{03}$ where $\Omega_{a}$ is the peak value of a $50$ ns square pulse,
and $\Omega_{b}$ is the Rabi frequency of a continuous wave laser.  Four-wave mixing condition ($\Delta k=0$) is assumed.\ The four
atomic levels are chosen as ($|0\rangle,$ $|1\rangle,$ $|2\rangle,$
$|3\rangle$) $=$ ($|$5S$_{1/2},$F=3$\rangle,$ $|5$P$_{3/2},$F=4$\rangle,$ $|4$D$_{5/2},$F=5$\rangle,$
$|5$P$_{3/2},$F=4$\rangle$). \ The natural decay
rate for atomic transition $|1\rangle\rightarrow|0\rangle$ or $|3\rangle
\rightarrow|0\rangle$ is $\gamma_{01}=\gamma_{03}=1/26$ ns and they have a
wavelength 780 nm. \ For atomic transition $|2\rangle\rightarrow|1\rangle$ or
$|2\rangle\rightarrow|3\rangle$ is $\gamma_{12}=\gamma_{32}=0.156\gamma_{03}$
\cite{gsgi} with a telecom wavelength 1.53$\mu$m. \ The scale factor of the
coupling constants for signal and idler transitions is $g_{s}/g_{i}=0.775.$

\begin{figure}
[ptb]
\begin{center}
\includegraphics[width=0.5\textwidth
]%
{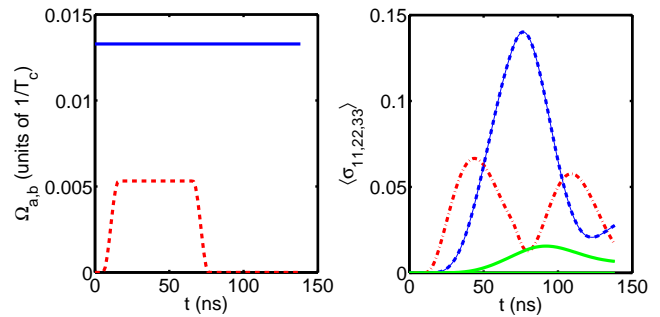}%
\caption{(Color online) Time-varying pump fields and time evolution of atomic populations.
(Left) The first pump field $\Omega_{a}$ (dotted-red) is a square pulse of
duration 50 ns and $\Omega_{b}$ is continuous wave (solid-blue). \ (Right)
The time evolution of the real part of populations for three atomic levels
$\sigma_{11}=\left\langle \tilde{\alpha}_{13}\right\rangle $ (dash dotted-red),
$\sigma_{22}=\left\langle \tilde{\alpha}_{12}\right\rangle $ (dotted-blue),
$\sigma_{33}=\left\langle \tilde{\alpha}_{11}\right\rangle $ (solid-green) at
$z=0,L$, and almost vanishing imaginary parts for all three of them. indicate
convergence of the ensemble averages.\ Note that these atomic populations are
uniform as a function of $z.$}%
\label{omega_population}%
\end{center}
\end{figure}

\begin{figure}
[ptb]
\begin{center}
\includegraphics[width=0.5\textwidth,height=0.35\textwidth
]%
{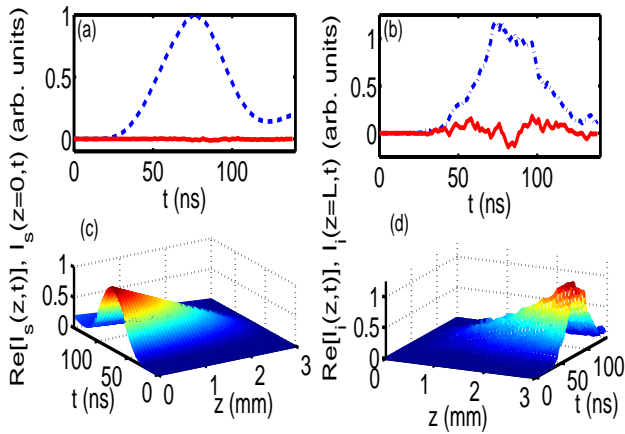}%
\caption{(Color online) Spatial-temporal intensity profiles of counter-propagating signal and idler
fields. \ (a) At $z=0,$ real (dashed-blue) and imaginary (solid-red) parts of
signal intensity. \ (b) At $z=L,$ real (dash dotted-blue) and imaginary
(solid-red) parts of idler intensity. \ (c) and (d) are spatial-temporal profiles for 
signal and idler intensities respectively. \ Both intensities are normalized by
the peak value of signal intensity that is $7.56\times10^{-12}$ $E_{c}^{2}$.
\ Note that the idler fluctuations and its non-vanishing imaginary part
indicate a relatively slower convergence compared with the signal intensity.
\ The ensemble size was 8$\times10^{5},$ and the atomic density $\rho=10^{10}%
$cm$^{-3}$.}%
\label{intensity}%
\end{center}
\end{figure}

\begin{figure}
[ptb]
\begin{center}
\includegraphics[width=0.5\textwidth,height=0.38\textwidth
]%
{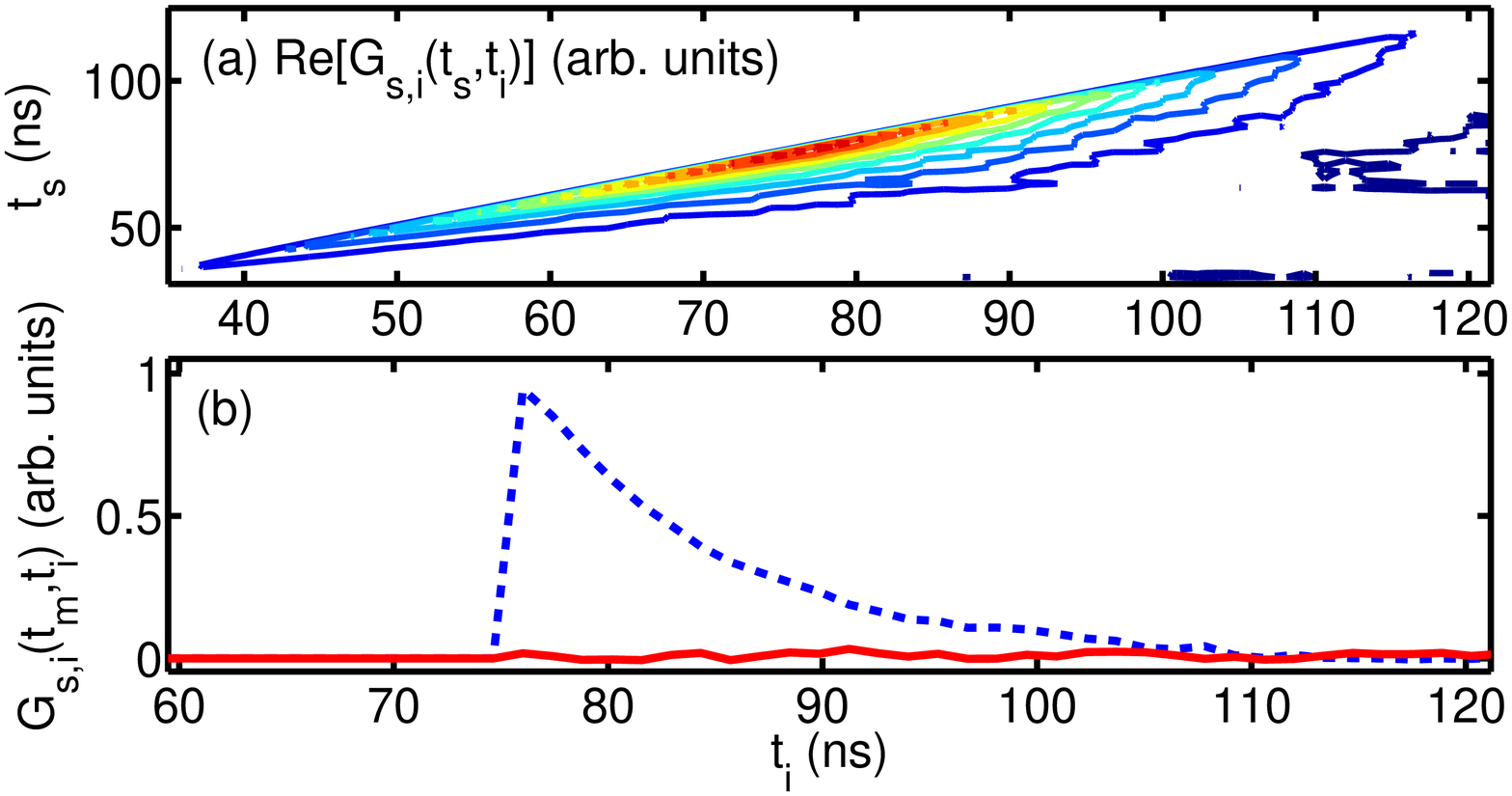}%
\caption{(Color online) Second-order correlation function $G_{s,i}(t_{s},t_{i}).$ The 2-D
contour plot of the real part of $G_{s,i}$ with a causal cut-off at
$t_{s}=t_{i}$ is shown in (a). \ The plot (b) gives a cross-section at
$t_{s}=t_{m}\approx75$ ns, which is normalized to the maximum of the real part
(dashed-blue) of $G_{s,i}.$ \ The imaginary part (solid-red) of $G_{s,i}$ is
nearly vanishing, and the number of realizations is 8$\times10^{5}$ for
$\rho=10^{10}$cm$^{-3}.$}%
\label{gsi_t}%
\end{center}
\end{figure}

We have investigated six different atomic densities from a dilute ensemble
with an optical density (opd) of 0.01 to a opd = 8.71. \ In Fig.\ref{omega_population}, \ref{intensity}, and \ref{gsi_t}, we take the atomic
density $\rho=10^{10}$ cm$^{-3}$ (opd = 2.18) for example, and the grid sizes
for dimensionless time $\Delta t=4$ and space $\Delta z=0.0007$ are chosen.
\ The convergence of the grid spacings is fixed in practice by convergence to
the signal intensity profile with an estimated relative error less than 0.5\%.%

The temporal profiles of the exciting lasers are shown in the left panel of
Fig.\ref{omega_population}. \ The atomic density is chosen as $\rho
=10^{10}$ cm$^{-3},$ and the cooperation time $T_{c}$ is 0.35 ns. \ The right
panel shows time evolution of atomic populations for levels $|1\rangle$,
$|2\rangle,$ and $|3\rangle$\ at $z=0,L,$ that are spatially uniform. \ The
populations are found by ensemble averaging the complex stochastic population
variables. \ The imaginary parts of the ensemble averages tend to zero as the
ensemble size is increased, and this is a useful indicator of convergence. \ In this example, the ensemble size was
8$\times10^{5}.$ \ The small rise after the pump pulse $\Omega_{a}$ is turned
off is due to the modulation caused by the pump pulse $\Omega_{b},$ which has
a generalized Rabi frequency $\sqrt{\Delta_{2}^{2}+4\Omega_{b}^{2}}$. This
influences also the intensity profiles and the correlation functions.%

In Fig.\ref{intensity}, we show counter-propagating signal ($-\hat{z}
$) and idler ($+\hat{z}$) field intensities at the respective ends of the atomic
ensemble and their spatial-temporal profiles respectively. \ The plots show the real and imaginary parts of the observables,
and both are normalized to the peak value of signal intensity. \ Note that the
characteristic field strength in terms of natural decay rate of the idler
transition ($\gamma_{03}$) and dipole moment ($d_{i}$) is $(d_{i}/\hbar
)E_{c}\approx36.3\gamma_{03}$. \ The fluctuation in the real idler field
intensity at $z=L$ and non-vanishing imaginary part indicates a slower
convergence compared to the signal field that has an almost vanishing
imaginary part. \ The slow convergence is a practical limitation of the
method. \ 

In Fig.\ref{gsi_t} (a), we show a contour plot of the second-order
correlation function $G_{s,i}(t_{s},t_{i})$ where $t_{i}\geq t_{s}.$ \ In
Figure \ref{gsi_t} (b), a section is shown through $t_{s}\approx75$ ns where
$G_{s,i}$ is at its maximum. \ The approximately exponential decay of
$G_{s,i}$ is clearly superradiant qualitatively consistent with the reference \cite{telecom}. \ The non-vanishing imaginary part of $G_{s,i}$
calculated by ensemble averaging is also shown in (b) and indicates a
reasonable convergence after 8$\times10^{5}$ realizations.  

In Table \ref{table1}, we display numerical parameters of our simulations for six
different atomic densities. \ The number of dimensions in space and time is
$M_{t}\times M_{z}$ with grid sizes ($\Delta t,\Delta z$) in terms of
cooperation time ($T_{c}$), length ($L_{c}$). \ The superradiant time scale
($T_{f}$) is found by fitting $G_{s,i}$ to an exponential function
($e^{-t/T_{f}}$), with $95\%$ confidence range.%

\begin{table}[h] \centering

\caption{Numerical simulation parameters for different atomic densities $\rho$.  Corresponding optical depth (opd),
time and space grids ($M_{t}\times M_{z}$) with grid sizes ($\Delta t,\Delta z$) in terms of cooperation time ($T_c$) and
length ($L_c$), and the fitted characteristic time $T_{f}$ for $G_{s,i}$ (see text).}%

\begin{tabular}
[c]{|c|c|c|c|c|c|}\cline{1-5}\cline{4-4}\cline{6-6}
$\rho($cm$^{-3})$ & opd & $M_{t}\times M_{z}$ &
\begin{tabular}
[c]{c}%
$\Delta t(T_{c}),$\\
$\Delta z(L_{c})$%
\end{tabular}
&
\begin{tabular}
[c]{c}%
$T_{c}($ns$),$\\
$L_{c}($m$)$%
\end{tabular}
&
\begin{tabular}
[c]{c}%
fitted $T_{f}$\\(ns)
\end{tabular}
\\\cline{1-5}\cline{4-4}\cline{6-6}
5$\times10^{7}$ & $0.01$ & $111\times42$ & 0.3, 5$\times10^{-5}$ & 4.89,
1.47 & $25.9$\\\hline
5$\times10^{8}$ & $0.11$ & $101\times44$ & 0.9, 1.5$\times10^{-4}$ & 1.55,
0.46 & $24.6$\\\hline
5$\times10^{9}$ & $1.09$ & $101\times42$ & 2.8, 4.5$\times10^{-4}$ & 0.49,
0.15 & $14.8$\\\hline
1$\times10^{10}$ & $2.18$ & $101\times42$ & 4.0, 7$\times10^{-4}$ & 0.35, 0.10 &
$9.4$\\\hline
2$\times10^{10}$ & $4.35$ & $101\times42$ & 5.5, 1$\times10^{-3}$ & 0.24,
0.07 & $5.0$\\\hline
4$\times10^{10}$ & $8.71$ & $101\times42$ & 8.0, 1.4$\times10^{-3}$ & 0.17,
0.06 & $3.1$\\\hline
\end{tabular}
\label{table1}%
\end{table}

\begin{figure}
[ptb]
\begin{center}
\includegraphics[width=0.52\textwidth,height=0.29\textwidth
]%
{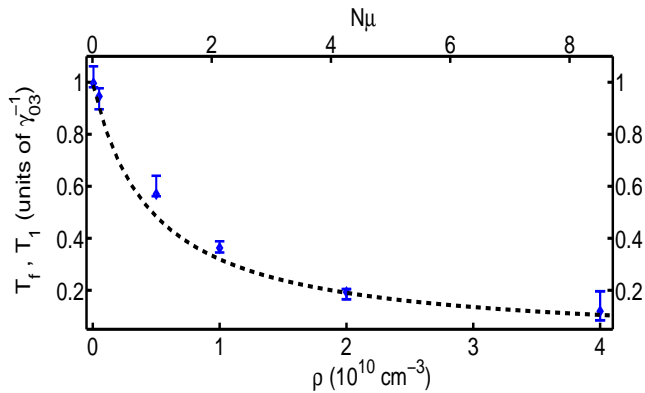}%
\caption{(Color online) Characteristic timescales, $T_{f}$ and $T_{1}$ vs atomic density
$\rho$ and the superradiant enhancement factor $N\mu$.
$\ T_{f}$ (dotted-blue) is the fitted characteristic timescale for
$G_{s,i}(t_{s}=t_{m},t_{i}=t_{m}+\tau)$ where $t_{m}$ is chosen at its
maximum, as in Figure \ref{gsi_t}. \ The error bars indicate the fitting
uncertainties. \ As a comparison, $T_{1}$=$\gamma_{03}^{-1}/$($N\mu+1$)
(dashed-black) is plotted where $\gamma_{03}^{-1}=26$ ns is the natural decay
time of D2 line of $^{85}$Rb atom, and $\mu$ is the geometrical constant for a
cylindrical atomic ensemble. \ The number of
realizations is 4$\times10^{5}$ for $\rho=5\times10^{7}$, $5\times10^{8}$, $5\times10^{9}$
cm$^{-3}$, 8$\times10^{5}$ for $\rho=10^{10}$, $2\times10^{10}$ cm$^{-3}$, and 16$\times10^{5}$ for $\rho=4\times 10^{10}$ cm$^{-3}$.}%
\label{gsi}%
\end{center}
\end{figure}

In Fig.\ref{gsi}, the characteristic time scale is plotted as a function of
atomic density and the factor $N\mu$, and shows faster decay for
optically denser atomic ensembles. \ We also plot the timescale $T_{1}%
=\gamma_{03}^{-1}/(N\mu+1)$ (ns) where $\mu$ is the geometrical constant for a cylindrical ensemble \cite{mu}. $\ $The natural decay time $\gamma_{03}^{-1}=26$ ns
corresponds to the D2 line of $^{85}$Rb. \ The error bar indicates the
deviation due to the fitting range from the peak of $G_{s,i}$ to approximately
25\% and 5\% of the peak value. \ The results of simulations are in good
qualitative agreement with the timescale of $T_{1}$ that can be regarded as a superradiant time constant of lower transition in a two-photon cascade \cite{QO:Scully, QL:Loudon}.  $T_{f}$ approaches independent atom behavior at lower
densities, which indicates no collective behavior as expected.  We note here that our simulations involve multiple excitations within the pumping condition similar to the experimental parameters \cite{telecom}.  The small deviation of $T_{f}$ and $T_{1}$ might be due to the multiple emissions considered in our simulations other than a two-photon source.  On the other hand the close asymptotic dependence of atomic density or optical depth in $T_{f}$ and $T_{1}$ indicates a strong correlation between signal and idler fields due to the four-wave mixing condition as required and crucial in experiment \cite{telecom}. 

For larger opd atomic ensembles, larger statistical ensembles are
necessary for numerical simulations to converge. \ The integration of
8$\times10^{5}$ realizations used in the case of $\rho=10^{10}$ cm$^{-3}$
consumes about 14 days with Matlab's parallel computing toolbox (function
"\textit{parfor"}) with a Dell precision workstation T7400 (64-bit Quad-Core
Intel Xeon processors).%


\section{Discussion and Conclusion}

The cascade atomic system studied here provides a source of telecommunication photons that are crucial for long distance quantum communication.  We may take advantage of such low loss transmission bandwidth in the DLCZ protocol for a quantum repeater.  The performance of the protocol relies on the efficiency of generating the cascade emission pair, which is better for a larger optical depth of the prepared atomic ensemble.  For other applications in quantum information science such as quantum swapping and quantum teleportation, the frequency space correlations also influence their success rates \cite{spectral}.  To utilize and implement the cascade emission in quantum communication, we characterize the emission properties, especially the signal-idler correlation function and its dependence on optical depths.  Its superradiant timescale indicates a broader spectral distribution which saturates the storage efficiency of idler pulse in an auxiliary atomic ensemble \cite{telecom} by means of EIT (electromagnetic induced transparency).  Therefore our calculation provides the minimal spectral window (1/$T_f$) of EIT to efficiently store and retrieve the idler pulse.

In summary, we have derived c-number Langevin equations in the positive-P representation
for the cascade signal-idler emission process in an atomic ensemble. \ The equations are solved numerically by a
stable and convergent semi-implicit difference method, while the
counter-propagating spatial evolution is solved by implementing the shooting
method. \ We investigate six different atomic densities readily obtainable in a
magneto-optical trap experiment. \ Signal and idler field intensities and
their correlation function are calculated by ensemble averages. \ Vanishing of
the unphysical imaginary parts within some tolerance is used as a guide to
convergence. \ We find an enhanced characteristic time scale for idler
emission in the second-order correlation functions from a dense atomic
ensemble, qualitatively consistent with the superradiance timescales used in a cylindrical dense atomic ensemble \cite{mu, telecom}.
\section*{ACKNOWLEDGMENTS}
We acknowledge support from NSF, USA and NSC, Taiwan, R. O. C., and thank T. A. B. Kennedy for guidance of this work.

\appendix
\section{Hamiltonian and Characteristic functions in Positive P-representation method}

The Hamiltonian $H$ is in Schr\"{o}dinger picture, and we separate it into two parts where $H_{0}$ is the free Hamiltonian of the atomic ensemble and one dimensional counter-propagating signal and idler fields, and $H_{I}$ is the interaction Hamiltonian of atoms interacting with two classical fields and two quantum fields (signal and idler) as shown in Fig.\ref{four}. \ Dipole approximation of $-\vec{d}\cdot\vec{E}$ and rotating wave approximation (RWA) have been made to these interactions. \ Using the standard quantization of electromagnetic field \cite{quantization}, we have%

\begin{align}
H_{0} & =\sum_{i=1}^{3}\sum_{l=-M}^{M}\hbar\omega_{i}\tilde{\sigma}_{ii}^{l}+\hbar\omega_{s}\sum_{l=-M}^{M}
\hat{a}_{s,l}^{\dag}\hat{a}_{s,l}\nonumber\\&+\hbar\sum_{l,l^{\prime}}\omega_{l^{\prime}l}\hat{a}_{s,l}^{\dag}\hat{a}%
_{s,l^{\prime}} +\hbar\omega_{i}\sum_{l=-M}^{M}\hat{a}_{i,l}^{\dag}\hat{a}_{i,l}\nonumber\\&+\hbar
\sum_{l,l^{\prime}}\omega_{ll^{\prime}}\hat{a}_{i,l}^{\dag}\hat{a}_{i,l^{\prime}}\text{ ,}\\
H_{I}  & =-\hbar\sum_{l=-M}^{M}\Big[\Omega_{a}(t)\tilde{\sigma}_{01}^{l\dagger}e^{ik_{a}z_{l}-i\omega_{a}t}
\nonumber\\&+\Omega_{b}(t)\tilde{\sigma}_{12}^{l\dagger}e^{-ik_{b}z_{l}-i\omega_{b}t}+h.c.\Big]\nonumber\\
& -\hbar\sum_{l=-M}^{M}\Big[  g_{s}\sqrt{2M+1}\tilde{\sigma}_{32}^{l\dagger}\hat{a}_{s,l}e^{-ik_{s}z_{l}} 
\nonumber\\&+g_{i}\sqrt{2M+1}\tilde{\sigma}_{03}^{l\dagger}\hat{a}_{i,l}e^{ik_{i}z_{l}}+h.c.\Big]
\end{align}
where $\tilde{\sigma}_{mn}^{l}\equiv\sum_{\mu}^{N_{z}}\hat{\sigma}_{mn}^{\mu,l}=\sum_{\mu}^{N_{z}}|m\rangle_{\mu}\langle n|\Big|_{r_{\mu}=z_{l}%
},~\Omega_{a}(t)\equiv f_{a}(t)d_{10}\mathcal{E}(k_{a})/(2\hbar),$ and $f_{a}$ is slow varying temporal profile without spatial dependence (ensemble scale
much less than pulse length). $g_{s}\equiv d_{23}\mathcal{E}(k_{s})/\hbar,~\mathcal{E}(k)=\sqrt{\hbar\omega/2\epsilon_{0}V}$ and $z_{m}%
=\frac{mL}{2M+1},~m=-M,...,M,$ and $L$ is the length of propagation that is equally split into $2M+1$ elements.  Commutation relations of field operators are $[\hat{a}_{l},\hat{a}_{l^{\prime}}^{\dag}]=\delta_{ll^{\prime}},$ and the matrix $\omega_{ll^{\prime}}\equiv\sum_{n}\frac{k_{n}c}{2M+1}e^{ik_{n}(z_{l}-z_{l^{\prime}})}$ 
accounts for field propagation by coupling the local mode operators where $k_n=2\pi n/L$. \ Note that the Rabi frequency is half of the standard definition.

The normally ordered exponential operator is chosen as%

\begin{align}
E(\lambda)  & =\prod_{l}E^{l}(\lambda),\nonumber\\
E^{l}(\lambda)  & =e^{i\lambda_{19}^{l}\tilde{\sigma}_{01}^{l\dagger}}e^{i\lambda_{18}^{l}\tilde{\sigma}_{12}^{l\dagger}}e^{i\lambda_{17}^{l}%
\tilde{\sigma}_{02}^{l\dagger}}e^{i\lambda_{16}^{l}\tilde{\sigma}_{13}^{l\dagger}}e^{i\lambda_{15}^{l}\tilde{\sigma}_{03}^{l\dagger}%
}e^{i\lambda_{14}^{l}\tilde{\sigma}_{32}^{l\dagger}}\times\nonumber\\&
e^{i\lambda_{13}^{l}\tilde{\sigma}_{11}^{l}}e^{i\lambda_{12}^{l}\tilde{\sigma}_{22}^{l}%
}e^{i\lambda_{11}^{l}\tilde{\sigma}_{33}^{l}}e^{i\lambda_{10}^{l}\tilde
{\sigma}_{32}^{l}} e^{i\lambda_{9}^{l}\tilde{\sigma}_{03}^{l}}e^{i\lambda_{8}^{l}\tilde{\sigma}_{13}^{l}}\times
\nonumber\\&e^{i\lambda_{7}^{l}\tilde{\sigma}_{02}^{l}}e^{i\lambda_{6}^{l}\tilde{\sigma}_{12}^{l}}e^{i\lambda_{5}^{l}\tilde{\sigma}_{01}^{l}%
}e^{i\lambda_{4}^{l}\hat{a}_{s,l}^{\dagger}}e^{i\lambda_{3}^{l}\hat{a}_{s,l}}e^{i\lambda_{2}^{l}\hat{a}_{i,l}^{\dagger}}e^{i\lambda_{1}^{l}\hat{a}_{i,l}%
}.\label{order}%
\end{align}

Aside from the atom-field interaction $\frac{\partial\rho}{\partial t}=\frac{1}{i\hbar}[H,\rho],$\ when dissipation from vacuum is considered
(single atomic decay), we can express them in terms of a Lindblad form where we have for the four-level atomic system,
\begin{align}
&\big(\frac{\partial\rho}{\partial t}\big)_{sp}=\nonumber\\&\sum_{l=-M}^{M}\sum_{\mu}^{N_{z}}\Big\{\frac{\gamma_{01}}{2}[2\hat{\sigma}_{01}^{\mu,l}\rho\hat
{\sigma}_{01}^{\mu,l\dagger}-\hat{\sigma}_{01}^{\mu,l\dagger}\hat{\sigma}_{01}^{\mu,l}\rho-\rho\hat{\sigma}_{01}^{\mu,l\dagger}\hat{\sigma}_{01}^{\mu,l}]
\nonumber\\& +\frac{\gamma_{12}}{2}[2\hat{\sigma}_{12}^{\mu,l}\rho\hat{\sigma}_{12}^{\mu,l\dagger}-\hat{\sigma}_{12}^{\mu,l\dagger}\hat{\sigma}_{12}^{\mu,l}%
\rho-\rho\hat{\sigma}_{12}^{\mu,l\dagger}\hat{\sigma}_{12}^{\mu,l}]\nonumber\\
& +\frac{\gamma_{32}}{2}[2\hat{\sigma}_{_{32}}^{\mu,l}\rho\hat{\sigma}_{_{32}}^{\mu,l\dagger}-\hat{\sigma}_{_{32}}^{\mu,l\dagger}\hat{\sigma}_{_{32}}%
^{\mu,l}\rho-\rho\hat{\sigma}_{_{32}}^{\mu,l\dagger}\hat{\sigma}_{_{32}}^{\mu,l}]\nonumber\\
& +\frac{\gamma_{03}}{2}[2\hat{\sigma}_{03}^{\mu,l}\rho\hat{\sigma}_{03}^{\mu,l\dagger}-\hat{\sigma}_{03}^{\mu,l\dagger}\hat{\sigma}_{03}^{\mu,l}%
\rho-\rho\hat{\sigma}_{03}^{\mu,l\dagger}\hat{\sigma}_{03}^{\mu,l}]\Big\}.
\end{align}

The characteristic functions can be calculated as

\begin{align}
\chi & =\text{Tr\{}E(\lambda)\rho\text{\},}\\
\frac{\partial\chi}{\partial t}  & =\text{Tr\{}E(\lambda)\frac{\partial\rho}{\partial t}\text{\}}\nonumber\\
&=\big(\frac{\partial\chi}{\partial t}\big)_{A}%
+\big(\frac{\partial\chi}{\partial t}\big)_{L} +\big(\frac{\partial\chi}{\partial t}\big)_{A-L}+\big(\frac{\partial\chi}{\partial t}\big)_{sp},\\
\big(\frac{\partial\chi}{\partial t}\big)_{A}  & =\text{Tr\{}E(\lambda)\frac{1}{i\hbar}[H_{A},\rho]\text{\}, }\nonumber\\
\big(\frac{\partial\chi}{\partial t}\big)_{L}& =\text{Tr\{}E(\lambda)\frac{1}{i\hbar}[H_{L},\rho]\text{\},}\nonumber\\
\big(\frac{\partial\chi}{\partial t}\big)_{A-L}  & =\text{Tr\{}E(\lambda)\frac{1}{i\hbar}[H_{A-L},\rho]\text{\}, }\nonumber\\
\big(\frac{\partial\chi}{\partial t}\big)_{sp}&=\text{Tr\{}E(\lambda)\big(\frac{\partial\rho}{\partial t}\big)_{sp}\text{\}}%
\end{align}
where $H_{0}=H_{A}+H_{L}$, $H_{A}$ is the atomic free evolution Hamiltonian, $H_{L}$ is the Hamiltonian for laser fields, and $H_{A-L}=H_{I}.$ 
The detail of derivations in various characteristic functions can be found in laser theory \cite{LT:Haken} or theory of light-atom interactions in atomic ensembles \cite{thesis}.

\section{Stochastic Differential Equation}

A distribution function can be found by Fourier transforming the characteristic functions,%

\begin{equation}
f(\vec{\alpha})=\frac{1}{(2\pi)^{n}}\int...\int e^{-i\vec{\alpha}\cdot\vec{\lambda}}\chi(\vec{\lambda})d\lambda_{1}...d\lambda_{n},
\end{equation}
then%

\begin{equation}
\frac{\partial f}{\partial t}=\frac{1}{(2\pi)^{n}}\int...\int e^{-i\vec{\alpha}\cdot\vec{\lambda}}\frac{\partial\chi}{\partial t}d\lambda_{1}...d\lambda_{n}.
\end{equation}

If $\frac{\partial\chi}{\partial t}=i\lambda_{\beta}\frac{\partial\chi}{\partial(i\lambda_{\gamma})}$, use integration by parts and neglect the
boundary terms, we have $\frac{\partial f}{\partial t}=-\frac{\partial}{\partial(\alpha_{\beta})}\alpha_{\gamma}f$ where a minus sign is from
$i\lambda_{\beta}$. \ Correspondingly, if $\frac{\partial\chi}{\partial t}=e^{i\lambda_{\beta}}$, we have \bigskip$\frac{\partial f}{\partial
t}=e^{-\frac{\partial}{\partial(\alpha_{\beta})}}$.

\subsection{Fokker-Planck equation}
Let
\begin{align}
&\frac{\partial f}{\partial t}=\mathcal{L}f=\sum_{l,l^{\prime}}\nonumber\\&
[\mathcal{L}_{A}\delta_{ll^{\prime}}+\mathcal{L}_{L}+\mathcal{L}_{A-L}^{(a)}%
\delta_{ll^{\prime}}+\mathcal{L}_{A-L}^{(b)}\delta_{ll^{\prime}}+\mathcal{L}_{sp}\delta_{ll^{\prime}}]f,
\end{align}
and we may neglect higher order derivatives (third order and
higher) in various $\mathcal{L}$'s. \ The validity of truncation to second order is due to
the expansion in the small parameter $1/N_{z}$. \ 

If the Fokker-Planck equation is
\begin{equation}
\frac{\partial f}{\partial t}=-\frac{\partial}{\partial\alpha}A_{\alpha
}f-\frac{\partial}{\partial\beta}A_{\beta}f+\frac{1}{2}(\frac{\partial^{2}%
}{\partial\alpha\partial\beta}+\frac{\partial^{2}}{\partial\beta\partial
\alpha})D_{\alpha\beta}f
\end{equation}
where $A$ and $D$ are drift and diffusion terms then we have a corresponding
classical Langevin equation%

\begin{equation}
\frac{\partial\alpha}{\partial t}=A_{\alpha}+\Gamma_{\alpha}\text{, }%
\frac{\partial\beta}{\partial t}=A_{\beta}+\Gamma_{\beta}%
\end{equation}
with a correlation function $\langle\Gamma_{\alpha}\Gamma_{\beta}%
\rangle=\delta(t-t^{\prime})D_{\alpha\beta}$. \ So now we can derive the equations of motion according to
various $\mathcal{L}$'s, but we postpone them and derivations of diffusion coefficients after the scaling is made for a dimensionless form in the next subsection.
The demonstration of various $\mathcal{L}$'s can be found in laser theory \cite{LT:Haken} or theory of light-atom interactions in atomic ensembles \cite{thesis}.

\subsection{Slowly varying envelopes and scaled equations of motion}

Here we introduce the slowly varying envelopes and define our cross-grained
collective atomic and field observables, then finally transform the equations
in a dimensionless form for later numerical simulations.  Define slow varying observables that
\begin{align}
& \widetilde{\alpha}_{5}(z,t)\equiv\frac{1}{N_{z}}\alpha_{5}^{l}%
e^{-ik_{a}z_{l}+i\omega_{a}t},~\widetilde{\alpha}_{6}(z,t)\equiv\frac{\alpha_{6}^{l}}%
{N_{z}}e^{ik_{b}z_{l}+i\omega_{b}t},\nonumber\\
& \widetilde{\alpha}_{7}(z,t)\equiv\frac{1}{N_{z}}\alpha_{7}^{l}%
e^{-ik_{a}z_{l}+ik_{b}z_{l}+i\omega_{b}t+i\omega_{a}t},\nonumber\\&~\widetilde{\alpha}%
_{8}(z,t)\equiv\frac{1}{N_{z}}\alpha_{8}^{l}e^{-i\omega_{a}t+i\omega
_{3}t+ik_{a}z_{l}-ik_{i}z_{l}},\nonumber
\end{align}\begin{align}
\text{ }  & \widetilde{\alpha}_{9}(z,t)\equiv\frac{1}{N_{z}}\alpha_{9}%
^{l}e^{-ik_{i}z_{l}+i\omega_{3}t},\widetilde{\alpha}_{11}(z,t)\equiv\frac
{1}{N_{z}}\alpha_{11}^{l},\nonumber\\
& \widetilde{\alpha}_{12}(z,t)\equiv\frac{1}{N_{z}}\alpha_{12}^{l}%
,\widetilde{\alpha}_{13}(z,t)\equiv\frac{1}{N_{z}}\alpha_{13}^{l},\nonumber\\
& \text{ }\widetilde{\alpha}_{14}(z,t)\equiv\frac{1}{N_{z}}\alpha_{14}%
^{l}e^{-i(\omega_{23}+\Delta_{2})t}e^{ik_{a}z_{l}-ik_{b}z_{l}-ik_{i}z_{l}}%
\end{align}
where $e^{i\Delta kz}=e^{ik_{a}z_{l}-ik_{b}z_{l}-ik_{i}z_{l}+ik_{s}z_{l}}$.  We note that

\begin{equation}
i\sum_{l^{\prime}}\omega_{ll^{\prime}}\alpha_{4}^{l^{\prime}}=c\frac{d}%
{dz_{l}}\alpha_{4}^{l}\text{, }-i\sum_{l^{\prime}}\omega_{ll^{\prime}}%
\alpha_{1}^{l^{\prime}}=-c\frac{\partial}{\partial z_{l}}\alpha_{1}^{l},
\end{equation}
and $\alpha_{0}^{l}=N_{z}-\alpha_{13}^{l}-\alpha_{12}^{l}-\alpha_{11}^{l}$, which will be used in later coupled equations.  Also for the field variables,%

\begin{align}
&E_{s}^{-}(z,t)\equiv\frac{g_{s}^{\ast}}{d_{i}/\hbar}\sqrt{2M+1}\alpha_{4}%
^{l}e^{-i\omega_{s}t},\nonumber\\&\text{ }E_{i}^{+}(z,t)\equiv\frac{g_{i}}{d_{i}/\hbar
}\sqrt{2M+1}\alpha_{1}^{l}e^{i\omega_{i}t},
\end{align}
where we use the idler dipole moment in signal field scaling for the purpose
of scale-free atomic equation of motions, so we need to keep in mind that in
calculating signal intensity or correlation function, an extra factor of
$(d_{i}/d_{s})^{2}$ needs to be taken care of.

We choose the central frequency of signal and idler as $\omega_{s}=\omega
_{23}+\Delta_{2},\omega_{i}=\omega_{3}$ where $\Delta_{1}=\omega_{a}%
-\omega_{1}$ and $\Delta_{2}=\omega_{a}+\omega_{b}-\omega_{2}$. \ With a
scaling of Arecchi-Courtens cooperation length \cite{scale}, we set up the
units of time, length, and field strength in the following,%

\begin{equation}
\text{ }L_{c}=cT_{c},\text{ }\frac{1}{T_{c}}=\sqrt{\frac{d_{i}^{2}n\omega_{i}}%
{2\hbar\epsilon_{0}}},\text{ }E_{c}=\frac{1}{T_{c}}\frac{1}{d_{i}/\hbar}.
\end{equation}

Now the slowly varying and dimensionless equations of motion with Langevin noises in Ito's form are%
\begin{widetext}
\begin{align}
\frac{\partial}{\partial t}\widetilde{\alpha}_{5}  & =(i\Delta_{1}%
-\frac{\gamma_{01}}{2})\widetilde{\alpha}_{5}+i\Omega_{a}(\widetilde{\alpha
}_{0}-\widetilde{\alpha}_{13})+i\Omega_{b}^{\ast}\widetilde{\alpha}%
_{7}-i\widetilde{\alpha}_{16}E_{i}^{+}+\mathcal{F}_{5},\nonumber\\
\frac{\partial}{\partial t}\widetilde{\alpha}_{6}  & =i(\Delta_{2}-\Delta
_{1}+i\frac{\gamma_{01}+\gamma_{2}}{2})\widetilde{\alpha}_{6}-i\Omega
_{a}^{\ast}\widetilde{\alpha}_{7}+i\Omega_{b}(\widetilde{\alpha}%
_{13}-\widetilde{\alpha}_{12})+i\widetilde{\alpha}_{8}E_{s}^{+}e^{-i\Delta
kz}+\mathcal{F}_{6},\nonumber\\
\frac{\partial}{\partial t}\widetilde{\alpha}_{7}  & =(i\Delta_{2}%
-\frac{\gamma_{2}}{2})\widetilde{\alpha}_{7}-i\Omega_{a}\widetilde{\alpha}%
_{6}+i\Omega_{b}\widetilde{\alpha}_{5}+i\widetilde{\alpha}_{9}E_{s}%
^{+}e^{-i\Delta kz}-i\widetilde{\alpha}_{10}E_{i}^{+}+\mathcal{F}%
_{7},\nonumber
\end{align}\begin{align}
\frac{\partial}{\partial t}\widetilde{\alpha}_{13}  & =-\gamma_{01}%
\widetilde{\alpha}_{13}+\gamma_{12}\widetilde{\alpha}_{12}+i\Omega
_{a}\widetilde{\alpha}_{19}-i\Omega_{a}^{\ast}\widetilde{\alpha}_{5}%
-i\Omega_{b}\widetilde{\alpha}_{18}+i\Omega_{b}^{\ast}\widetilde{\alpha}%
_{6}+\mathcal{F}_{13},\nonumber\\
\frac{\partial}{\partial t}\widetilde{\alpha}_{12}  & =-\gamma_{2}%
\widetilde{\alpha}_{12}+i\Omega_{b}\widetilde{\alpha}_{18}-i\Omega_{b}^{\ast
}\widetilde{\alpha}_{6}+i\widetilde{\alpha}_{14}E_{s}^{+}e^{-i\Delta
kz}-i\widetilde{\alpha}_{10}E_{s}^{-}e^{i\Delta kz}+\mathcal{F}_{12}%
,\nonumber\\
\frac{\partial}{\partial t}\widetilde{\alpha}_{11}  & =-\gamma_{03}%
\widetilde{\alpha}_{11}+\gamma_{32}\widetilde{\alpha}_{12}-i\widetilde{\alpha
}_{14}E_{s}^{+}e^{-i\Delta kz}+i\widetilde{\alpha}_{10}E_{s}^{-}e^{i\Delta
kz}+i\widetilde{\alpha}_{15}E_{i}^{+}-i\widetilde{\alpha}_{9}E_{i}%
^{-}+\mathcal{F}_{11},\nonumber\\
\frac{\partial}{\partial t}\widetilde{\alpha}_{8}  & =-(i\Delta_{1}%
+\frac{\gamma_{01}+\gamma_{03}}{2})\widetilde{\alpha}_{8}-i\Omega_{a}^{\ast
}\widetilde{\alpha}_{9}-i\Omega_{b}\widetilde{\alpha}_{14}+i\widetilde{\alpha
}_{6}E_{s}^{-}e^{i\Delta kz}+i\widetilde{\alpha}_{19}E_{i}^{+}+\mathcal{F}%
_{8},\nonumber\\
\frac{\partial}{\partial t}\widetilde{\alpha}_{9}  & =-\frac{\gamma_{03}}%
{2}\widetilde{\alpha}_{9}-i\Omega_{a}\widetilde{\alpha}_{8}+i\widetilde
{\alpha}_{7}E_{s}^{-}e^{i\Delta kz}+i(\widetilde{\alpha}_{0}-\widetilde
{\alpha}_{11})E_{i}^{+}+\mathcal{F}_{9},\nonumber\\
\frac{\partial}{\partial t}\widetilde{\alpha}_{14}  & =-(i\Delta_{2}%
+\frac{\gamma_{03}+\gamma_{2}}{2})\widetilde{\alpha}_{14}-i\Omega_{b}^{\ast
}\widetilde{\alpha}_{8}+i(\widetilde{\alpha}_{12}-\widetilde{\alpha}%
_{11})E_{s}^{-}e^{i\Delta kz}+i\widetilde{\alpha}_{17}E_{i}^{+}+\mathcal{F}%
_{14},
& \label{bloch2}%
\end{align}\end{widetext}
where $\gamma_{2}=\gamma_{12}+\gamma_{32}$, and field propagation equations are%
\begin{align}
(\frac{\partial}{\partial t}-\frac{\partial}{\partial z})E_{s}^{-}  &
=-i\widetilde{\alpha}_{14}e^{-i\Delta kz}\frac{|g_{s}|^{2}}{|g_{i}|^{2}%
}+\mathcal{F}_{4},\nonumber\\
(\frac{\partial}{\partial t}+\frac{\partial}{\partial z})E_{i}^{+}  &
=i\widetilde{\alpha}_{9}+\mathcal{F}_{1},
\end{align}

where $\frac{|g_{s}|^{2}}{|g_{i}|^{2}}$ is a unit transformation factor from
the signal field strength to the idler one. \ For a recognizable format of the
above equations used in the main context, we change the labels in the below,%

\begin{align}
& \widetilde{\alpha}_{5}\leftrightarrow\pi_{01},\text{ }\widetilde{\alpha}%
_{6}\leftrightarrow\pi_{12},\text{ }\widetilde{\alpha}_{7}\leftrightarrow
\pi_{02},\text{ }\widetilde{\alpha}_{8}\leftrightarrow\pi_{13},\text{
}\widetilde{\alpha}_{9}\leftrightarrow\pi_{03},\nonumber\\&\text{ }\widetilde{\alpha}%
_{10}\leftrightarrow\pi_{32},\text{ }\widetilde{\alpha}_{11}\leftrightarrow
\pi_{33},\widetilde{\alpha}_{12}\leftrightarrow\pi_{22},\text{ }\widetilde{\alpha
}_{13}\leftrightarrow\pi_{11},\nonumber\\&\text{ }\widetilde{\alpha}_{14}\leftrightarrow
\pi_{32}^{\dag},\text{ }\widetilde{\alpha}_{15}\leftrightarrow\pi_{03}^{\dag
},\text{ }\widetilde{\alpha}_{16}\leftrightarrow\pi_{13}^{\dag},\text{
}\widetilde{\alpha}_{17}\leftrightarrow\pi_{02}^{\dag},\nonumber\\&\text{ } \widetilde{\alpha}_{18}\leftrightarrow\pi_{12}^{\dag},\text{ }%
\widetilde{\alpha}_{19}\leftrightarrow\pi_{01}^{\dag},
\end{align}
where $\pi_{ij}$ is the stochastic variable that corresponds to the atomic
populations of state $|i\rangle$ when $i=j$ and to atomic coherence when
$i\neq j$. \ Note that the associated c-number Langevin noises are changed accordingly.

The Langevin noises are defined as%

\begin{align}
\mathcal{F}_{5}(z,t)  & =\frac{\Gamma_{5}^{l}}{N_{z}}e^{-ik_{a}z_{l}%
+i\omega_{a}t},\mathcal{F}_{6}(z,t)=\frac{\Gamma_{6}^{l}}{N_{z}}%
e^{ik_{b}z_{l}+i\omega_{b}t},\nonumber\\
\mathcal{F}_{7}(z,t)  & =\frac{\Gamma_{7}^{l}}{N_{z}}e^{-ik_{a}z_{l}%
+ik_{b}z_{l}+i\omega_{b}t+i\omega_{a}t},\mathcal{F}_{13}(z,t)=\frac{\Gamma_{13}^{l}}{N_{z}%
},\nonumber\\\mathcal{F}_{11}(z,t)  & =\frac{\Gamma_{11}^{l}}{N_{z}},\mathcal{F}%
_{8}(z,t)=\frac{1}{N_{z}}\Gamma_{8}^{l}e^{-i\omega_{a}t+i\omega_{3}%
t+ik_{a}z_{l}-ik_{i}z_{l}},\nonumber\\\mathcal{F}_{14}(z,t)&=\frac{1}{N_{z}}\Gamma_{14}^{l}%
e^{-i(\omega_{23}+\Delta_{2})t}e^{ik_{a}z_{l}-ik_{b}z_{l}-ik_{i}z_{l}%
},\nonumber\\\mathcal{F}_{9}(z,t)   &=\frac{\Gamma_{9}^{l}}{N_{z}}e^{-ik_{i}z_{l}+i\omega_{3}t},\mathcal{F}_{12}(z,t)=\frac{\Gamma_{12}^{l}}{N_{z}},\nonumber\\
\mathcal{F}_{4}(z,t)  & =\frac{g_{s}^{\ast}}{d_{i}/\hbar}\sqrt{2M+1}%
e^{-i\omega_{s}t}\Gamma_{4}^{l},\nonumber\\\mathcal{F}_{1}(z,t)&=\frac{g_{i}}{d_{i}/\hbar
}\sqrt{2M+1}e^{i\omega_{i}t}\Gamma_{1}^{l}%
\end{align}
where other Langevin noises can be found by using the correspondence, for example, $\mathcal{F}_{5}^{\ast}\leftrightarrow
\mathcal{F}_{19}$.

Before we proceed to formulate the diffusion coefficients, we need to be
careful about the scaling factor for the transformation to continuous
variables when numerical simulation is applied. \ Take $\left\langle
\mathcal{F}_{6}\mathcal{F}_{5}\right\rangle $ for example,%
\begin{align}
& \left\langle \mathcal{F}_{6}(z,t)\mathcal{F}_{5}(z^{\prime},t^{\prime
})\right\rangle \nonumber\\
& =\frac{1}{N_{z}^{2}}e^{ik_{b}z_{l}+i\omega_{b}t}e^{-ik_{a}z_{l^{\prime}%
}+i\omega_{a}t^{\prime}}\left\langle \Gamma_{6}^{l}\Gamma_{5}^{l^{\prime}%
}\right\rangle \nonumber\\
& =\frac{1}{N_{z}^{2}}e^{ik_{b}z_{l}+i\omega_{b}t}e^{-ik_{a}z_{l}+i\omega
_{a}t}[i\Omega_{a}e^{ik_{a}z_{l}-i\omega_{a}t}\alpha_{6}^{l}\nonumber\\&+ig_{i}\sqrt
{2M+1}e^{ik_{i}z_{l}}\alpha_{10}^{l}\alpha_{1}^{l}]\delta(t-t^{\prime}%
)\delta_{ll^{\prime}}\nonumber\\
& =\frac{1}{N_{c}}\left[  i(\Omega_{a}T_{c})\widetilde{\alpha}_{6}%
+i\widetilde{\alpha}_{10}(E_{i}^{+}/E_{c})\right]  \frac{1}{T_{c}^{2}}%
\delta(t-t^{\prime})T_{c}\times\nonumber\\&\delta(z-z^{\prime})L_{c}%
\end{align}
where we have used $\lim_{M\rightarrow\infty}\frac{2M+1}{L}\delta_{ll^{\prime
}}=\delta(z-z^{\prime})$, $2M+1=\frac{N}{N_{z}},$ and $N_{c}=\frac{NL_{c}}{L}$
is the cooperation number. \ Then we have the dimensionless form of diffusion coefficients.%

\begin{align}
T_{c}^{2}\left\langle \mathcal{F}_{6}(\tilde{z},\tilde{t})\mathcal{F}_{5}(\tilde{z}^{\prime}%
,\tilde{t}^{\prime})\right\rangle  & =\frac{D_{6,5}}{N_{c}}\delta(\tilde{t}-\tilde{t}^{\prime}%
)\delta(\tilde{z}-\tilde{z}^{\prime})\\
D_{6,5}  & =\left[  i\Omega_{a}\widetilde{\alpha}_{6}+i\widetilde{\alpha}%
_{10}E_{i}^{+}\right]  .
\end{align}

The dimensionless diffusion coefficients $D_{ij}$ are%
\begin{widetext}
\begin{align}
(\text{i})D_{5,5}  & =-i2\Omega_{a}\widetilde{\alpha}_{5};\text{ }%
D_{5,6}=i(\Omega_{a}\widetilde{\alpha}_{6}+\widetilde{\alpha}_{10}E_{i}%
^{+});\text{ }D_{5,7}=-i\Omega_{a}\widetilde{\alpha}_{7};\text{ }
D_{5,8}   =i(\Omega_{a}\widetilde{\alpha}_{8}+(\widetilde{\alpha}%
_{11}-\widetilde{\alpha}_{13})E_{i}^{+});\text{ }\nonumber\\D_{5,9}&=-i(\Omega
_{a}\widetilde{\alpha}_{9}+\widetilde{\alpha}_{5}E_{i}^{+});\text{
}D_{5,11}  =-i\widetilde{\alpha}_{16}E_{i}^{+};\text{ }D_{5,13}%
=i\widetilde{\alpha}_{16}E_{i}^{+};\text{ }D_{5,14}=-i\widetilde{\alpha}%
_{18}E_{i}^{+};\text{ }D_{5,19}=\gamma_{12}\widetilde{\alpha}_{12};\nonumber\\
(\text{ii})D_{6,6}  & =-i2\Omega_{b}\widetilde{\alpha}_{6};\text{ }%
D_{6,8}=-i\Omega_{b}\widetilde{\alpha}_{8};\text{ }D_{6,10}=-i\Omega
_{b}\widetilde{\alpha}_{10};\text{ }
D_{6,13}   =-i\Omega_{a}^{\ast}\widetilde{\alpha}_{7}+\gamma_{01}%
\widetilde{\alpha}_{6};\text{ }\nonumber\\D_{6,16}&=-i\widetilde{\alpha}_{7}E_{i}%
^{-}+\gamma_{01}\widetilde{\alpha}_{10};\text{ }D_{6,18}=\gamma_{01}%
\widetilde{\alpha}_{12};\nonumber\\
(\text{iii})D_{7,8}  & =-i\widetilde{\alpha}_{6}E_{i}^{+};\text{ }%
D_{7,9}=-i\widetilde{\alpha}_{7}E_{i}^{+};\nonumber\\
(\text{iv})D_{8,9}  & =-i\widetilde{\alpha}_{8}E_{i}^{+};\text{ }%
D_{8,10}=i\Omega_{b}(\widetilde{\alpha}_{12}-\widetilde{\alpha}_{11});\text{
}D_{8,11}=i\Omega_{b}\widetilde{\alpha}_{14};\text{ }
D_{8,12}  =-i\Omega_{b}\widetilde{\alpha}_{14};\nonumber\\D_{8,13}& =-i\Omega_{a}^{\ast
}\widetilde{\alpha}_{9}+i\widetilde{\alpha}_{19}E_{i}^{+}+\gamma
_{01}\widetilde{\alpha}_{8};\text{ }
D_{8,16}   =i\widetilde{\alpha}_{15}E_{i}^{+}-i\widetilde{\alpha}_{9}%
E_{i}^{-}+\gamma_{01}\widetilde{\alpha}_{11}+\gamma_{32}\widetilde{\alpha
}_{12};\text{ }D_{8,18}=i\widetilde{\alpha}_{17}E_{i}^{+}+\gamma
_{01}\widetilde{\alpha}_{14};\nonumber\\
(\text{v})D_{9,9}  & =-i2\widetilde{\alpha}_{9}E_{i}^{+};\text{ }%
D_{9,10}=i\widetilde{\alpha}_{10}E_{i}^{+};\text{ }D_{9,15}=\gamma
_{32}\widetilde{\alpha}_{12};\nonumber\\
(\text{vi})D_{10,10}  & =-i2\widetilde{\alpha}_{10}E_{s}^{+}e^{-i\Delta
kz};\text{ }D_{10,11}=i(\Omega_{b}\widetilde{\alpha}_{16}-\widetilde{\alpha
}_{7}E_{i}^{-})+\gamma_{03}\widetilde{\alpha}_{10};\text{ }
D_{10,13}   =-i\Omega_{b}\widetilde{\alpha}_{16};\nonumber\\D_{10,14}&=i\Omega
_{b}\widetilde{\alpha}_{18}-i\Omega_{b}^{\ast}\widetilde{\alpha}_{6}%
+\gamma_{03}\widetilde{\alpha}_{12};\text{ }D_{10,19}=i\widetilde{\alpha}%
_{6}E_{i}^{-};\nonumber\\
(\text{vii})D_{11,11}  & =i\widetilde{\alpha}_{14}E_{s}^{+}e^{-i\Delta
kz}-i\widetilde{\alpha}_{10}E_{s}^{-}e^{i\Delta kz}+i\widetilde{\alpha}%
_{15}E_{i}^{+}-i\widetilde{\alpha}_{9}E_{i}^{-}+\gamma_{32}\widetilde{\alpha
}_{12}+\gamma_{03}\widetilde{\alpha}_{11};\nonumber\\
D_{11,12}  & =i\widetilde{\alpha}_{10}E_{s}^{-}e^{i\Delta kz}-i\widetilde
{\alpha}_{14}E_{s}^{+}e^{-i\Delta kz}-\gamma_{32}\widetilde{\alpha}%
_{12};\nonumber\\
(\text{viii})D_{12,12}  & =i\Omega_{b}\widetilde{\alpha}_{18}-i\Omega
_{b}^{\ast}\widetilde{\alpha}_{6}-i\widetilde{\alpha}_{10}E_{s}^{-}e^{i\Delta
kz}+i\widetilde{\alpha}_{14}E_{s}^{+}e^{-i\Delta kz}+\gamma_{2}\widetilde
{\alpha}_{12};
D_{12,13}  =-i\Omega_{b}\widetilde{\alpha}_{18}+i\Omega_{b}^{\ast}%
\widetilde{\alpha}_{6}-\gamma_{12}\widetilde{\alpha}_{12};\nonumber\\
(\text{ix})D_{13,13}  & =i\Omega_{a}\widetilde{\alpha}_{19}-i\Omega_{a}^{\ast
}\widetilde{\alpha}_{5}+i\Omega_{b}\widetilde{\alpha}_{18}-i\Omega_{b}^{\ast
}\widetilde{\alpha}_{6}+\gamma_{01}\widetilde{\alpha}_{13}+\gamma
_{12}\widetilde{\alpha}_{12};\nonumber\\
(\text{x})D_{3,8}  & =\frac{|g_{s}|^{2}}{|g_{i}|^{2}}i\widetilde{\alpha}%
_{6}e^{i\Delta kz};\text{ }D_{3,9}=\frac{|g_{s}|^{2}}{|g_{i}|^{2}}%
i\widetilde{\alpha}_{7}e^{i\Delta kz}.
\end{align}
\end{widetext}

Before going further to set up the stochastic differential equation in the next subsection, we remark on the alternative method to derive the diffusion coefficients from the Heisenberg-Langevin
approach with Einstein relations \cite{LP:Sargent, QO:Scully, Fleischhauer94}, and it provides the important check for
Fokker-Planck equations.  We note here that a symmetric property of the
diffusion coefficients is within Fokker-Planck equation, whereas the quantum
diffusion coefficients in quantum Langevin equation do not have symmetric
property simply because the quantum operators do not necessarily commute with
each other. 

\subsection{Ito and Stratonovich stochastic differential equations}

The c-number Langevin equations derived from Fokker-Planck equations have a
direct correspondence to Ito-type stochastic differential equations \cite{QN:Gardiner, SM:Gardiner}. 
\ In stochastic simulations, it is important to find the expressions of Langevin
noises\ from diffusion coefficients.

For any symmetric diffusion matrix $D(\alpha)$, it can always be factorized into%

\begin{equation}
D(\alpha)=B(\alpha)B^{T}(\alpha)
\end{equation}

where $B$ $\rightarrow$ $BS$ (an orthogonal matrix$S$ that $SS^{T}=I$)
preserves the diffusion matrix so $B$ is not unique. \ The matrix $B$ is in
terms of the Langevin noises where $\xi_{i}dt=dW_{t}^{i}$ (Wiener process) and
$\left\langle \xi_{i}(t)\xi_{j}(t^{\prime})\right\rangle =\delta_{ij}%
\delta(t-t^{\prime})$ and the $\xi_{i}$ below is just a random number in
Gaussian distribution with zero mean and unit variance.

In numerical simulation, we use the semi-implicit algorithm that guarantees
the stability and convergence in the integration of stochastic differential
equations. \ So a transformation from Ito to Stratonovich-type stochastic
differential equation is necessary,%

\begin{align}
dx_{t}^{i}  & =A_{i}(t,\overrightarrow{x_{t}})dt+\sum\limits_{j}%
B_{ij}(t,\overrightarrow{x_{t}})dW_{t}^{j}\text{ \ (Ito)}\\
dx_{t}^{i}  & =[A_{i}(t,\overrightarrow{x_{t}})-\frac{1}{2}\sum\limits_{j}%
\sum\limits_{k}B_{jk}(t,\overrightarrow{x_{t}})\frac{\partial}{\partial x^{j}%
}B_{ik}(t,\overrightarrow{x_{t}})]dt\nonumber\\
& +\sum\limits_{j}B_{ij}(t,\overrightarrow{x_{t}})dW_{t}^{j}\text{
\ (Stratonovich)}%
\end{align}
where a correction in drift term appears due to the transformation.

In the end we have the full equations with 19 variables in the positive-P
representation, 64 diffusion matrix elements, and 117 noise terms (random
number generators). \ Nonvanishing corrections in drift terms are only for $\widetilde{\alpha}_{5}$, $\widetilde{\alpha}_{6}$, 
$\widetilde{\alpha}_{9}$, $\widetilde{\alpha}_{10}$, $\widetilde{\alpha}_{11}$, $\widetilde{\alpha}_{12}$, $\widetilde{\alpha}_{13}$, and 
they are $i\Omega_{a}/2$, $i\Omega_{b}$, $iE_{i}^{+}$, $iE_{s}^{+}/2$, $(-3\gamma_{03}+\gamma_{32})/4$, $-\gamma_{2}/4$, $(-5\gamma_{01}+\gamma_{12})/4$ respectively.

The Langevin noises can be formulated as a non-square form \cite{QO:Walls,Smith1}, and in numerical simulations, we have a factor $\frac{1}{\sqrt{N_{c}\Delta t\Delta z}}$ for Langevin noises $\mathcal{F}$ and $\frac{1}{N_{c}\Delta t\Delta z}$
for correction terms.

\end{document}